\documentclass[namedreferences]{solarphysics}

\usepackage[hyperref,optionalrh,showbiblabels]{spr-sola-addons} 
\usepackage{graphicx}        
\usepackage{color}           
\usepackage{breakurl}        
\usepackage{xcolor}

\chardef\us=`\_

\newcommand{\kms}{{km s$^{-1}$}}
\usepackage{gensymb}
\graphicspath{{./}{figs/}}
\usepackage{hyperref}
\hypersetup{
    colorlinks=true,
    linkcolor=blue,
    citecolor=blue,
    filecolor=blue,      
    urlcolor=blue,
}

\begin{document}

\begin{article}
\begin{opening}

\title{Comparative Analysis of the Solar Wind: Modeling Charge State Distributions in the Heliosphere}

\author[addressref={aff1,aff2},corref,email={md4469@nyu.edu}]{\inits{M.}\fnm{Marcus}~\lnm{Dupont}\orcid{0000-0003-3356-880X}}
\author[addressref=aff2,email={chengcaishen@cfa.harvard.edu}]{\inits{C.}\fnm{Chengcai}~\lnm{Shen}\orcid{0000-0002-9258-4490}}
\author[addressref=aff2,email={namurphy@cfa.harvard.edu}]{\inits{N.A.}\fnm{Nicholas}~\lnm{Murphy}\orcid{0000-0001-6628-8033}}

\address[id=aff1]{Center for Cosmology and Particle Physics, New York University\\
                    726 Broadway\\
                    New York, NY, 10009, USA
                    }
\address[id=aff2]{Center for Astrophysics $\vert$ Harvard \& Smithsonian\\
60 Garden St. \\
Cambridge, MA 02138, USA}

\runningauthor{Dupont et al.}
\runningtitle{Comparative Analysis of the Solar Wind}

\begin{abstract}
Non-equilibrium ionization (NEI) is a key process often times
neglected when modeling astrophysical plasmas with
thermodynamical timescales
much shorter than the timescales for ionization and recombination. 
In this paper, we perform NEI modeling on a magnetohydrodynamic (MHD) simulation of the solar wind during the Whole Sun Month (Carrington Rotation 1913 from 1996 August 22 to September 18), and compare the resulting charge state distributions with in situ measurements made with the Solar Wind Ion Composition Spectrometer (SWICS) on Ulysses.
We trace the wind trajectory back to 20$R_\odot$ using the velocity measured by Ulysses at about 4 AU, and obtain the plasma flow trajectory within 20$R_\odot$ using the MHD simulation data performed by Predictive Science Inc. We assume that the wind started from the solar surface and was initially in ionization equilibrium, and analyze the time-dependent ionization state of solar wind along each of the trajectories.
In our analysis we: (1) obtain charge state densities and ratios for slow and fast winds based on the NEI model, and compare them with in situ observations; and (2) measure the ``freeze-in'' distance for several ions observed by SWICS to determine a possible correlation between when the ionization states become fixed and 
the electron density and outward velocity of the plasma at the freeze-in height.  This study provides a stringent test on comparing charge state distributions from the outer corona predicted from simulations with in situ measurements made in the far heliosphere. This showcases the challenges in matching plasma conditions near the corona to those observed in interplanetary space.
\end{abstract}
\end{opening}

\section{Introduction} \label{sec:intro}
The solar wind is a continuous flow of charged particles emanating from the Sun. The charge states of solar wind plasma provide diagnostic opportunities that can help unravel the mechanisms behind the solar wind acceleration, heating, and origin. Spacecraft such as the Advanced Composition Explorer (ACE), Ulysses \citep{1986ASSL..123..173G}, Hinode \citep{2006cosp...36.3642T}, and the Solar and Heliospheric Observatory \citep[SOHO;][]{1995SSRv...72...81D} have aided in collecting charge state information to better constrain these complex distribution mechanisms. The continuous data provided by these spacecraft have helped researchers divide the solar wind into two main components, the slow wind ($\sim 300$ \kms) and the fast wind ($\sim 750$ \kms). Newer spacecraft such as the Parker Solar Probe \citep{2016SSRv..204....7F} and the Solar Orbiter \citep{2013SoPh..285...25M} offer modern technologies with incredibly complex instruments that provide an even broader picture of the Sun's features, especially with the advantages of in situ measurement. In situ charge observations provide sharp constraints on solar wind models.  This abundance of observation has driven various theoretical models that seek to explain the solar wind acceleration and heating. Non-equilibrium ionization (NEI) modeling of solar wind plasma is a key strategy for providing constraints on these models. 

Time-dependent (or non-equilibrium) ionization becomes relevant when the thermodynamic state of the plasma changes faster than the typical timescales for ionization and recombination. In low-density regions in the solar wind, the plasma is expected to be out of equilibrium because the ionization/recombination time-scales become significantly longer than the dynamic time-scale of plasma. Moreover, the outward moving plasma in the solar wind does not have enough time to reach ionization equilibrium once the density drops significantly below the dense coronal atmosphere at the solar surface. Because of this, the ion species at ``freeze-in'' in the solar wind can be used to infer the temperature history during solar wind formation as well as place constraints on the solar wind source regions \citep{1979ApJ...229L.101D, 1997SoPh..171..345K,1981BAAS...13..812O, 2000ApJ...532L..71E}. Time-dependent ionization is important for interpreting both remote sensing and in situ observations (e.g., \citealt{2007ApJ...667..602R, 2012ApJ...761...48L}). Accounting for non-equilibrium ionization can have a significant difference on the interpretation of the observed structure of the solar atmosphere.
By observing spectroscopic line and continuum intensities, one is typically able to use time-dependent ionization as a diagnostic tool to 
determine the plasma properties and analyse the relevant heating/cooling mechanisms of the plasma. For example, \citet{2016ApJ...817...46M} presented an analysis of the non-equilibrium properties of O$^{3+}$ and Si$^{3+}$ by combining 2D MHD simulations and IRIS's spectroscopic observations. They looked at the Si IV/O IV line intensities in the quiet Sun and flaring active regions, and found that they were able to reproduce the observed line ratios when they fully considered the non-equilibrium case.
Additionally, \citet{2017SPD....4810615K}  
investigated the heating processes of coronal loops in the soft X-ray band to constrain the heating/cooling timescales of hot coronal lines. The emission measure of the coronal loop in the soft X-ray spectra is used to estimate the temperature timescale due to the fact that soft X-ray spectral lines typically correspond to highly ionized particles; and, as a result,
\citet{2017SPD....4810615K} presented that the line intensities showed a departure from ionization equilibrium where the degree of departure varied depending on the heat inputs for the coronal loop.

In general, the solar wind is not in ionization equilibrium. The electron temperature therefore does not match the local ionic charge state of the solar wind. The fast solar wind is believed to come from the solar coronal holes \citep{1973SoPh...29..505K}, but pinpointing the source of slower, highly-ionized wind regions remains an active area of research \citep{Abbo:2016}. Current theories speculate that they may come from coronal holes with diverging magnetic fields \citep{Wang:1992, Cranmer:2009}, the tips of helmet streamers \citep{Einaudi:1999, Lapenta:2005}, or from interchange reconnection near coronal hole boundaries \citep{Fisk:2001, Antiochos:2011}. A key way to go about exploring the source of the slow solar wind is by examining the inner conditions of the corona through in situ measurements of heavy ions present \citep{1968ApJ...152L...3H,1983ApJ...275..354O, 2007SSRv..130..139G}. Previous studies worked at this by modeling various magnetic field geometries (open or closed) and solar explosions to account for constituents present in the solar wind \citep{2011ApJ...730..103G,2012ApJ...761...48L, 2012ApJ...758L..21L,2015ApJ...806...55O,2017SoPh..292...90R}. The series of articles produced
by \citet{2012ApJ...761...48L, 2012ApJ...758L..21L, 2014ApJ...790..111L} provide a summary of plasma conditions as the plasma traverses the transition region outward into the fast solar wind. They found that, while the transition region lines were well predicted, the fast wind ionic charge states were underpredicted, indicating that the fast wind streams did not have time to collisionally ionize before departure. They also tested the models of \citet{1995JGR...10021577H}, \citet{2007ApJS..171..520C}, and \citet{2013ApJ...778..176O}. The models were found to be in better agreement with observation when taking the source of the fast solar wind to be from the corona as opposed to the chromosphere. These types of benchmarks are useful for developing and improving upon current theories on the coronal heating and solar wind acceleration problems.

In this paper we  analyze the charge state evolution of solar wind plasma that was observed during the Whole Sun Month (WSM, 1996 10 August to 8 September). The WSM was a coordinated observing campaign during Carrington Rotation (CR) 1913 from 1996 August 22 to Sep 18 to characterize and model the solar corona and wind around solar minimum \citep{Galvin:1999, Riley:1999}.  Solar wind models are heavily dependent on boundary conditions --- that is, boundaries involving the magnetic field in the lower corona and conditions surrounding the heliosphere \citep{2008SSRv..136..565A}.
We invoked a well developed solar wind model produced by the \textit{Magnetohydrodynamics Around a Sphere (MAS)} code for the WSM \citep{1999JGR...104.9809L, 2009ApJ...690..902L} to generate solar wind conditions. We applied the simulated solar wind parameters (temperature, density, velocity, and angular position) to a non-equilibrium ionization code entitled \textit{NEI} in order to extract ionic information from the stream traces such as the evolution of various charge states and abundance ratios to compare with the quantities observed by the Solar Wind Ion Composition Spectrometer \citep[SWICS;][]{1983ESASP1050...75G, 1992A&AS...92..267G} \citep{1992A&AS...92..267G} aboard Ulysses. We analyzed the time-dependent ionization state of plasma until it becomes frozen-in
beyond the corona wherein we compare these ionization states
with those measured by the SWICS instrument. 
This approach can ultimately be used to constrain the heating processes of the solar wind by identifying
bulk charge states and narrowing down their lower boundary heating mechanisms.
Note that with current magnetogram data only being collected from approximately the Earth viewing angle, there is restriction on the observational points which is a known limitation of MHD simulations. Such restrictions can introduce systematic errors
in solar wind simulations, but CR 1913 occurred near solar minimum 
which allowed for a simpler magnetic configuration in the MAS model.


We also focus on the steady outflows above pseudo-streamers, loop-like structures that separate coronal holes of the same polarity \citep{2014ApJ...787L...3R}, and relevant NEI effects on the solar wind 
composition. While MHD models predict accurate wind speeds that motivate the study of turbulence in the heliosphere 
\citep{1990PhyD...46..177C, 1997NPGeo...4..185H, 1999GeoRL..26.1801S,2003JGRA..108.1012P, 2015ApJ...814L..19W, 2017AGUFMSH33B2771A}, they also allow one to address the fast/slow wind interaction regions 
\citep{2018arXiv180609604K, 2018LRSP...15....1R}. 

This paper is organized as follows. In Section \ref{sec:data},
we describe the acquisition and utilization of the Ulysses data and discuss the MAS model for the Whole Sun Month
interval. In Section \ref{sec:method}, we compare the NEI calculations
to SWICS data. In Section \ref{sec:results} we present our results.
In Section \ref{sec:discussion}, we conclude with a discussion of
the relevance of our work.

\section{Data Preparation} \label{sec:data}
The WSM interval was a cumulative modeling and data effort which sought to understand large-scale structures that may persist for several rotations in the solar corona  \citep{Galvin:1999}. In situ observations contributed to this campaign by providing three-dimensional data out into the interplanetary medium through the observations of spacecraft such as Ulysses.
The data we are most interested in 
are the trace elemental and ionic charge compositions measured by Ulysses/SWICS\@.
During 1996 August 22 to September 18, the Ulysses spacecraft moves from 28$\degree$ to 26$\degree$ heliographic latitude at 4.25 AU from the solar center in the heliographic coordinate system. 
The main channel SWICS used to analyze and identify the solar wind ions utilized a combination of a time-of-flight sensor, a post-acceleration detector, electrostatic deflection, and residual energy measurement.
SWICS was uniquely able to measure both the elemental and ionic charge distributions
alongside the mean wind speeds of all of the major solar wind ions (H, He, C, N, O, Ne, Mg, S, Si, and Fe). In our analysis, we assume the abundances of \citet{Schmelz:2012} to calculate theoretical charge states for comparison with SWICS measurements. As mentioned before, measuring the heavy ions in the solar wind is indicative of 
the plasma conditions in the very early evolution (inner corona) of the solar wind as it
travels out into interplanetary space. Moreover, during the WSM campaign Ulysses began sampling increasingly variable wind, a characteristic of the bands of solar wind variability at lower and higher latitudes \citep{1997GeoRL..24..309G}. Having components like the kinetic temperature, mean speeds, and ionic distributions of the heavy ions in the wind gives rise to the underlying physics happening at the solar corona.

We used HelioPy \citep{heliopy_0_5_1} to acquire the SWICS data and extract information including the spacecraft coordinates, solar wind velocity and density, and abundance ratios for multiple ions. Once the 
coordinates of the spacecraft were acquired, we then matched
the observation position with the coordinates 
in the MHD simulation by assuming the kinematics of the 
plasma were entirely radial. We used the maximum time resolution (3 hr) provided by SWICS for the CR 1913 period for our analysis. This included more than 200 data points starting from September 3 to September 29 (accounting for the 11 days the plasma takes to go from 20 $R_\odot$ to Ulysses/SWICS) of detailed constituents of the solar wind at a distance of 4.25 AU\@. We used all available ions provided by the SWICS database, which included O$^{6+}$, C$^{6+}$, Ne$^{8+}$, Mg$^{10+}$, Si$^{9+}$, S$^{10+}$, and Fe$^{11+}$.
The above plasma and ion properties are used to compared with the following model simulation based on NEI calculation. 

A known source of error is that the magnetogram data used as a boundary condition for the far side of the Sun in this simulation is roughly two weeks old.  This is particularly relevant because Ulysses was on the opposite side of the Sun from Earth during this period.  Comparing the charge states at Ulysses during this interval with those predicted from NEI modeling of the simulation provides a stringent test of the accuracy of this simulation. \citet{Riley:1999} performed a similar analysis of comparing charge states between the Sun and Ulysses using two mapping techniques. They compared the results from the commonly used ballistic wind speed approximation with those computed from a two-dimensional inverse MHD algorithm and found that their MHD backwards mapping technique compared more closely to observation. With processes such as rarefaction and compression being possible beyond our MHD modeling boundary, a constant speed approximation might not capture the fast and slow wind stream interactions happening before Ulysses/SWICS samples the wind. Differences between the predicted and observed charge states provide information on how simulations of the solar wind can be improved.

We perform the NEI analysis based on the well-developed coronal 
MAS model from Predictive Science Inc.\ (PSI). 
We analyze a simulation of the Whole Sun Month (WSM) interval that overlapped with CR 1913 from 1996 August 22 to September 18, and obtain the primary plasma variables (velocity, temperature, and density) to apply the NEI calculation.
The earlier 
version of this MHD model has been successfully used to compare the 
white light polarization emission to Mauna Loa Solar Observatory MK3 images
\citep{1999JGR...104.9809L}, and also used to study the pseudo-streamer
and helmet streamer during CR 2067 \citep{2015SoPh..290.2043A}. A newer
model has been reported by \citet{2009ApJ...690..902L}, in which they used
this model with an improved energy equation to investigate different heating
terms in WSM model. The energy equation in the model contains parameterized
coronal heating, parallel thermal conduction, radiative losses, and the
acceleration of solar wind by Alfv\'en waves \citep{2009ApJ...690..902L}. The
synoptic 
magnetic field map for the WSM period is obtained from
the MDI instrument on SOHO\@. They made quantitative comparisons with emission
in EUV from the EIT instrument on SOHO, and with X-ray emission from the soft
X-ray telescope on Yohkoh. 
In this work, we use the updated model with higher spatial resolution than the
\citet{2009ApJ...690..902L} model, with $221\times275\times 431$ grid points, which
correspondingly included more small-scale structures in the photospheric magnetic
field. This improved MAS model also has been used to analyze NEI properties of a
pseudo-streamer during CR 1913 and compare with SOHO/UVCS observations in detail
\citep{2017ApJ...850...26S}. In this work, we are interested in the solar wind properties at higher altitudes above this well-studied pseudo-streamer in \citep{2017ApJ...850...26S}, and focus on freeze-in conditions along different trajectories. 

\section{Method of NEI analysis} \label{sec:method}
\subsection{Eigenvalue Method}\label{sec: eigenvalue}
\begin{figure*}[htp]
\includegraphics[width=\textwidth]{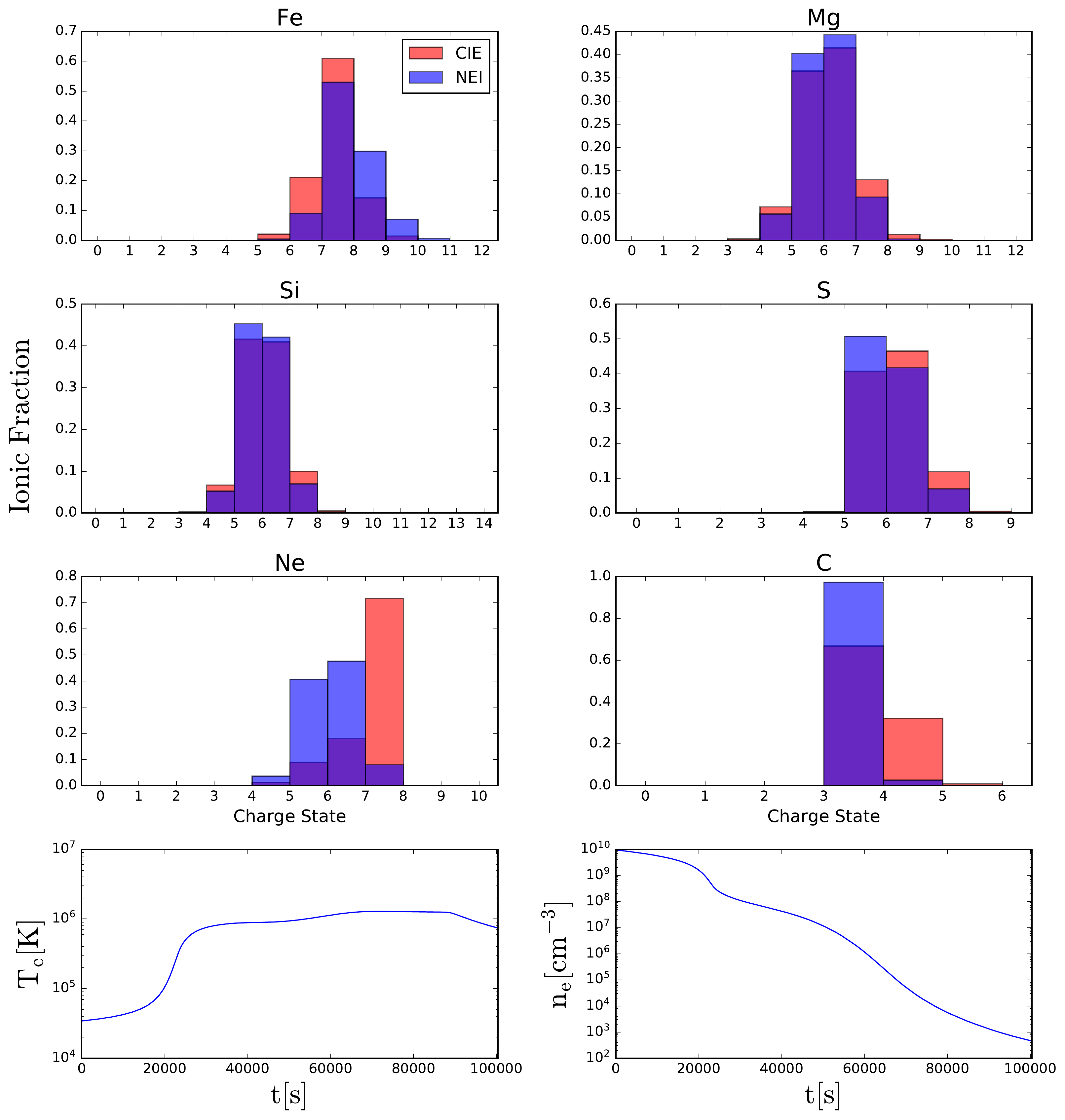}
\caption{Shown in these figures are the differences in ion population between a single sample stream trace for the equilibrium (red) and non-equilibrium (blue) cases. 
These are the modeled charge states at
20 $R_\odot$ and with a final velocity of 526 \kms. 
}
\label{fig: bar_plots}
\end{figure*}
The ionic charge state of plasma depends on its evolutionary history. We solve the time-dependent ionization equations in Lagrangian form to obtain the charge state by using the Eigenvalue method
\citep{1984Ap&SS..98..367M, 1985ApJ...291..544H, 2015A&C....12....1S} accompanied by an adaptive time-step that elongates or compresses the time-step depending 
on plasma conditions in order to retain computational efficiency.
The rate of change of the ion population fraction $X_i$ for a specific ion with integer charge $i$ is
\begin{equation}\label{eq: nei_equation}
\frac{dX_i}{dt} = n_e(t)(I_{i-1}X_{i-1} + R_iX_{i+1} - I_iX_i - R_{i-1}X_i).
\end{equation}
Here,
$n_e$ is the electron number density.
The total ionization and recombination rate coefficients for this ion are $I_i$ and $R_i$, respectively.
These rates were obtained from the CHIANTI version 8.0.7 atomic database
\citep{1997A&AS..125..149D, 2013ApJ...763...86L,
DelZann_2015A&A...582A..56D,
2016JPhB...49g4009Y}. An example of the key differences between collisional ionization equilibrium (CIE) and NEI is shown in Figure (\ref{fig: bar_plots}).
The ionization and recombination rates are functions of the plasma temperature, and do not strongly depend on the density. In a rapidly evolving system where the plasma density and temperature vary with time, the time-dependent ionization equations become non-linear.
In individual cases with constant temperature, the evolution of ionic fraction $X_i$
tends to its equilibrium state corresponding to this particular temperature. In more general cases, a small time-step is required to accurately capture changes of $X_i$ with time. Therefore, we follow the Eigenvalue method to solve time-dependent ionization equations to benefit from its robust and efficient numerical scheme. In summary, the Eigenvalue method takes Eq.\ (\ref{eq: nei_equation}) and formulates it into a matrix eigenvalue equation. The key matrix is diagonalizable and comprises all of the ionization and recombination rates for the respective ion. The updated ionization state vector is computed after diagonalizing the matrix of rates and extracting the eigenvalues and matrix of eigenvectors. Mathematically this is formulated as,
\begin{equation}\label{eq: matrix_eq}
\frac{d\textbf{F}}{dt} = n_e(\hat{\textbf{A}} \cdot \textbf{F}),
\end{equation}
where $\textbf{F}$ is the vector containing the ion fractions 
composed with $Z + 1$ ionization levels for an element of atomic number $Z$,
and $\hat{\textbf{A}}$ is the matrix containing the ionization and
recombination rate coefficients as shown in Eq.\ (\ref{eq: nei_equation}).
Once one diagonalizes the coefficient matrix $\hat{\textbf{A}}$,
the solution of Eq.\ (\ref{eq: matrix_eq}) can be written as:
\begin{equation}\label{eq: eigen_eq}
\frac{d\textbf{F}^\prime}{dt} = n_e\hat{\lambda} \cdot \textbf{F}^\prime,
\end{equation}
where $\hat{\lambda}$ is a diagonal matrix containing the eigenvalues calculated from
the initial coefficient matrix $\hat{\textbf{A}}$, and
$\textbf{F}^\prime \equiv \hat{\textbf{V}}^{-1}\textbf{F}$ is the new vector formed by multiplying both sides with a matrix of eigenvectors $\hat{\textbf{V}}$ with eigenvalues $\hat{\lambda}$. In a situation where the plasma temperature is constant and the corresponding ionization and 
recombination rates are also constant, Eq.\ (\ref{eq: eigen_eq})
yields the elementary exponential decay solution: $\textbf{F}^\prime = \textbf{F}_0$ exp($-n_e\hat{\lambda} t$).
In this case, \textbf{F}$_0$ is the initial ion fraction 
vector for the element in question. The robustness and speed
of this method allowed us to compute the ionization balance quite
easily and speedily as compared to other numerical schemes like
the Runge-Kutta algorithm. 
Note that this equation is stable even under conditions of
very large time-steps which would cause the system to go into a 
state of equilibrium. 
We then solve Eq. (\ref{eq: nei_equation}) using an open source Python project  {\it NEI-modeling/NEI\footnote{https://github.com/NEI-modeling/NEI}} and using eigenvalue tables created based on the method reported in 
\citet{2015A&C....12....1S}. 

\subsection{Getting the Streamtraces from MAS}\label{sec: streams}

In order to solve above ionization equations in a Lagrangian framework, we first trace plasma blobs in the Eulerian framework of MHD simulations, from the solar surface to the top boundary of the MHD model. The coordinate system the Ulysses spacecraft used was the
\textit{Heliocentric Radial, Tangential, and Normal (HGRTN)}
system \citep{2002P&SS...50..217F} which didn't 
require any major coordinate transformations 
in this work  because the  Heliographic Carrington coordinates match with those used in the MAS model. Since the spacecraft
was at a distance of approximately 4 AU and the upper boundary of this MHD model is 20 $R_\odot$ from the solar center, we accounted for the time delay
of the plasma blob leaving 20 $R_\odot$ to reach Ulysses by first averaging
out an observed velocity of the instrument from 1996 August 1 to October 1, and then used the mean velocity throughout
the WSM interval. This assumption allowed us to arrive at
a time delay of 11 days. 
Due to the Ulysses spacecraft having such a long period of orbit relative to the time-frame of this analysis, we treated it as stationary throughout CR 1913. In doing so, we assumed the bulk outflow speeds measured were purely from the radial component of the MAS model velocity. This assumption was motivated by the velocity contour map produced by the MAS model (see Figure \ref{fig: fieldlines}).
As an effect, we shifted the Ulysses observations
11 days after the WSM interval to allow time for the plasma
from CR 1913 to reach the SWICS instrument. Once these 
shifted dates and coordinates were passed into the MAS model at 20 $R_\odot$, it outputted a stream trace corresponding to each Ulysses observational data point.
Below 20 $R_\odot$, we use ADVECT, a program developed by PSI, to trace the plasma movement using time-dependent velocity fields obtained from the MAS model, and interpolate temperature and density in space and time along each of the tracing trajectories.
Thus each stream trace possesses a unique history of
temperature, density, velocity, and acceleration. 

\subsection{Performing NEI Calculations}\label{sec: nei}

Once the stream trace data was collected from the MAS model, 
it was a matter of folding the stream properties through
the NEI models. As aforementioned, each stream line 
possessed various values of temperature, density,
and velocity, so an evolutionary charge state 
distribution from the NEI model had to be
calculated for each individual trace. Since the SWICS
instrument only tracked the abundant elements in the solar wind
--- Fe, O, C, Ne, Mg, S, and Si --- \citep{1992A&AS...92..267G},
we performed non-equilibrium ionization calculations for all of these 
elements. 
The following NEI calculations start from 1.02 $R_\odot$ where the plasma number density is usually as high as $10^{10}$ cm$^{-3}$ and the plasma is assumed to be in equilibrium ionization states.
This allowed us to solve the time-dependent ionization equations and track the charge state 
distributions for each element present within the plasma traveling up
each flux tube out to 20 $R_\odot$.
The final charge states
measured at the end of the simulations were stored for each of the
streamers to later be compared to the in situ measurements.

\begin{figure*}
\includegraphics[width=\textwidth]{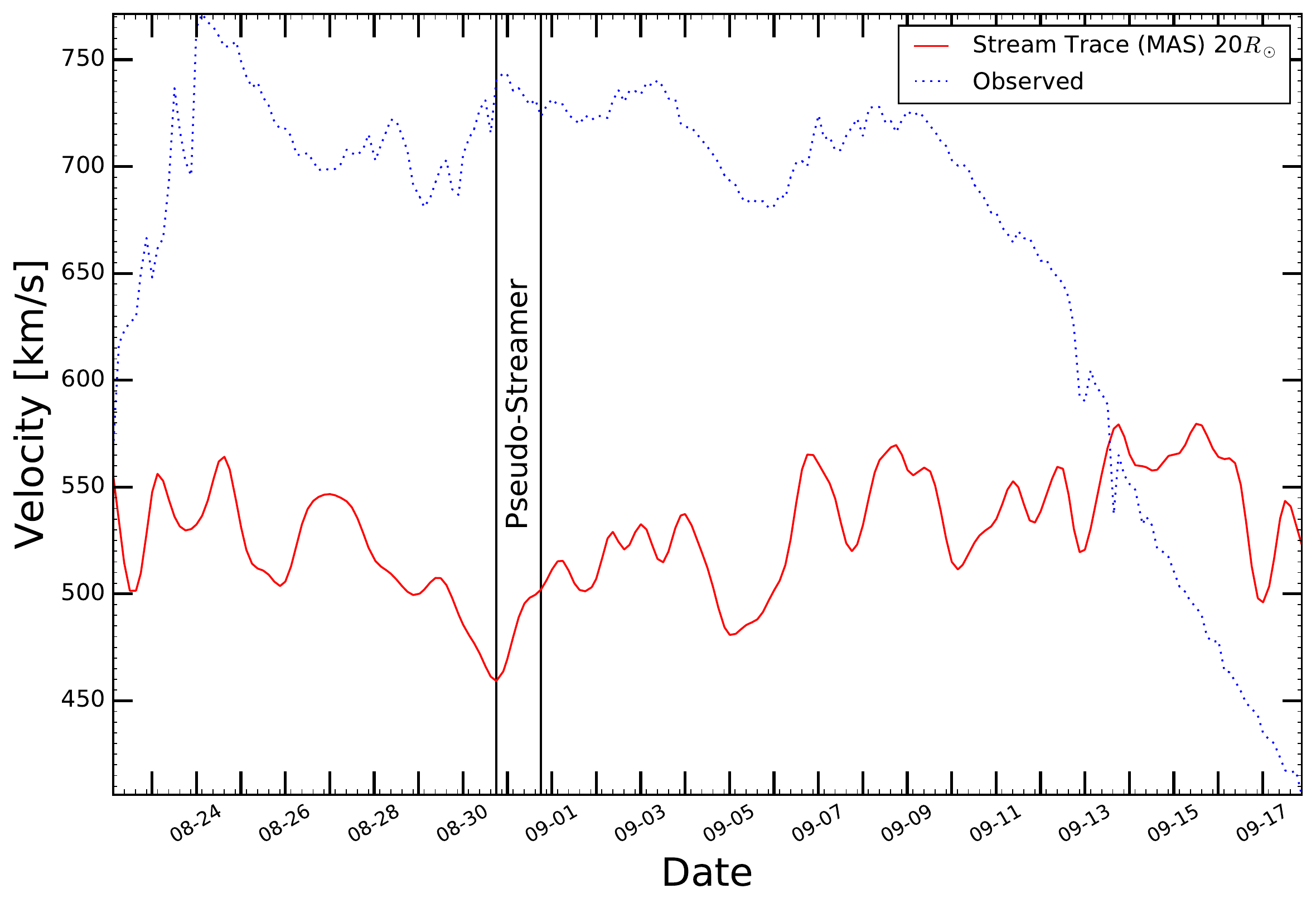}
\caption{The compared velocities between the MAS model at 20 $R_\odot$ and 
the mean velocities observed by the Ulysses/SWICS instrument during the
Whole Sun Month rotation (CR 1913). The rectangular bar indicates when the outflow
produced by the pseudo-streamer would have subtended the observational zone of SWICS.}
\label{fig: velocity}
\end{figure*}

\section{Results} \label{sec:results}

Calculating the final charge states meant that we could 
compare them within situ observations when the charge state is frozen-in already.
This meant that, to good approximation, the final
plasma state calculated at 20 $R_\odot$
consists of the same charge state distributions as expected in the solar wind 
measured in the outer heliosphere by 
SWICS\@. After the calculating the ionization
population for each stream trace, we compared the frozen-in
plasma with the relative densities (to O$^{6+}$) of Si$^{9+}$, S$^{10+}$,
Mg$^{10+}$, Ne$^{8+}$, Fe$^{8+}$, and C$^{6+}$.
We also compared the O$^{7+}$/O$^{6+}$ and C$^{6+}$/C$^{5+}$
abundance ratios with those found on SWICS\@. 

\subsection{Velocity Comparison}\label{sec: v_compare}
\begin{figure*}
\includegraphics[width=0.439\textwidth]{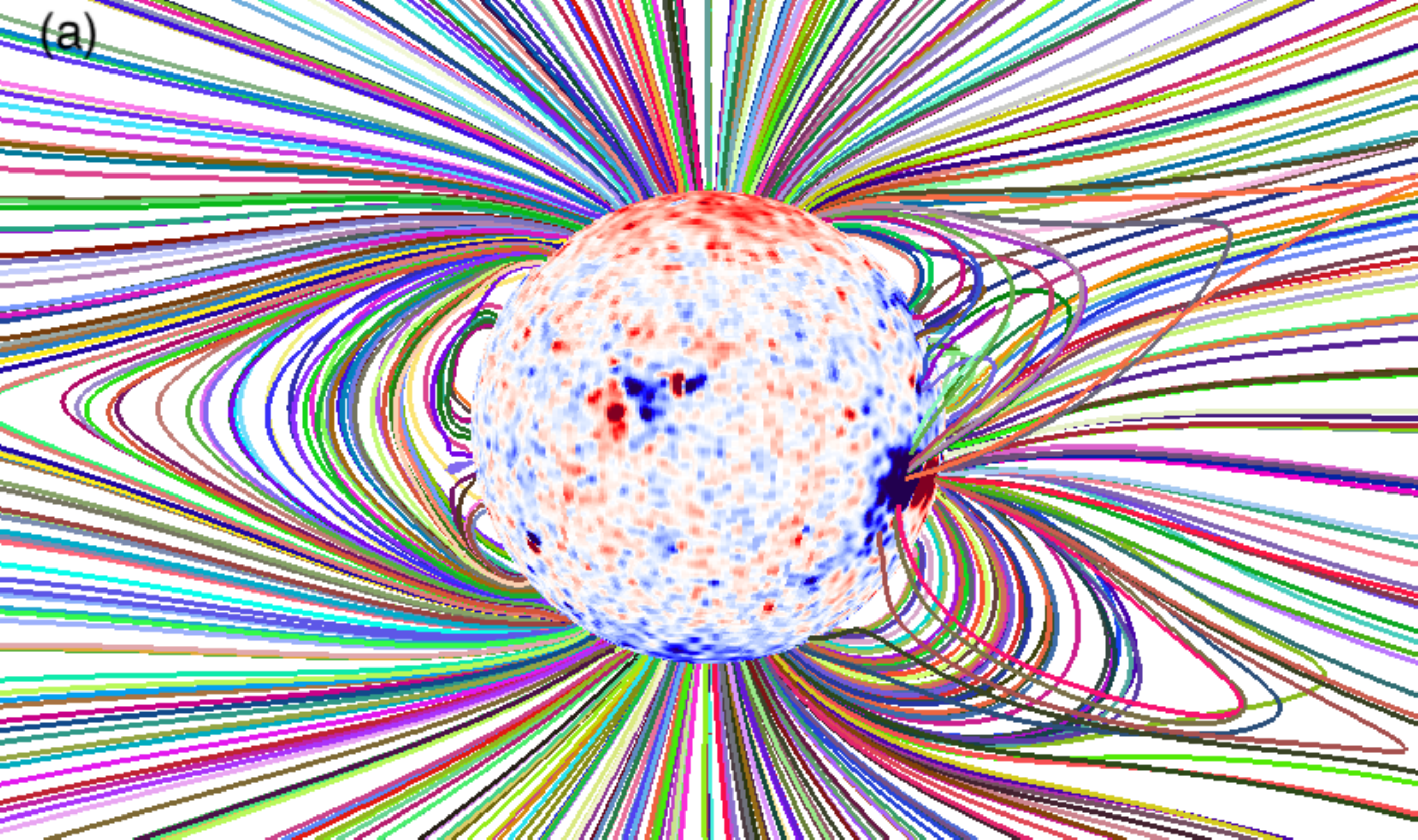}
\hfill
\includegraphics[width=0.55\textwidth]{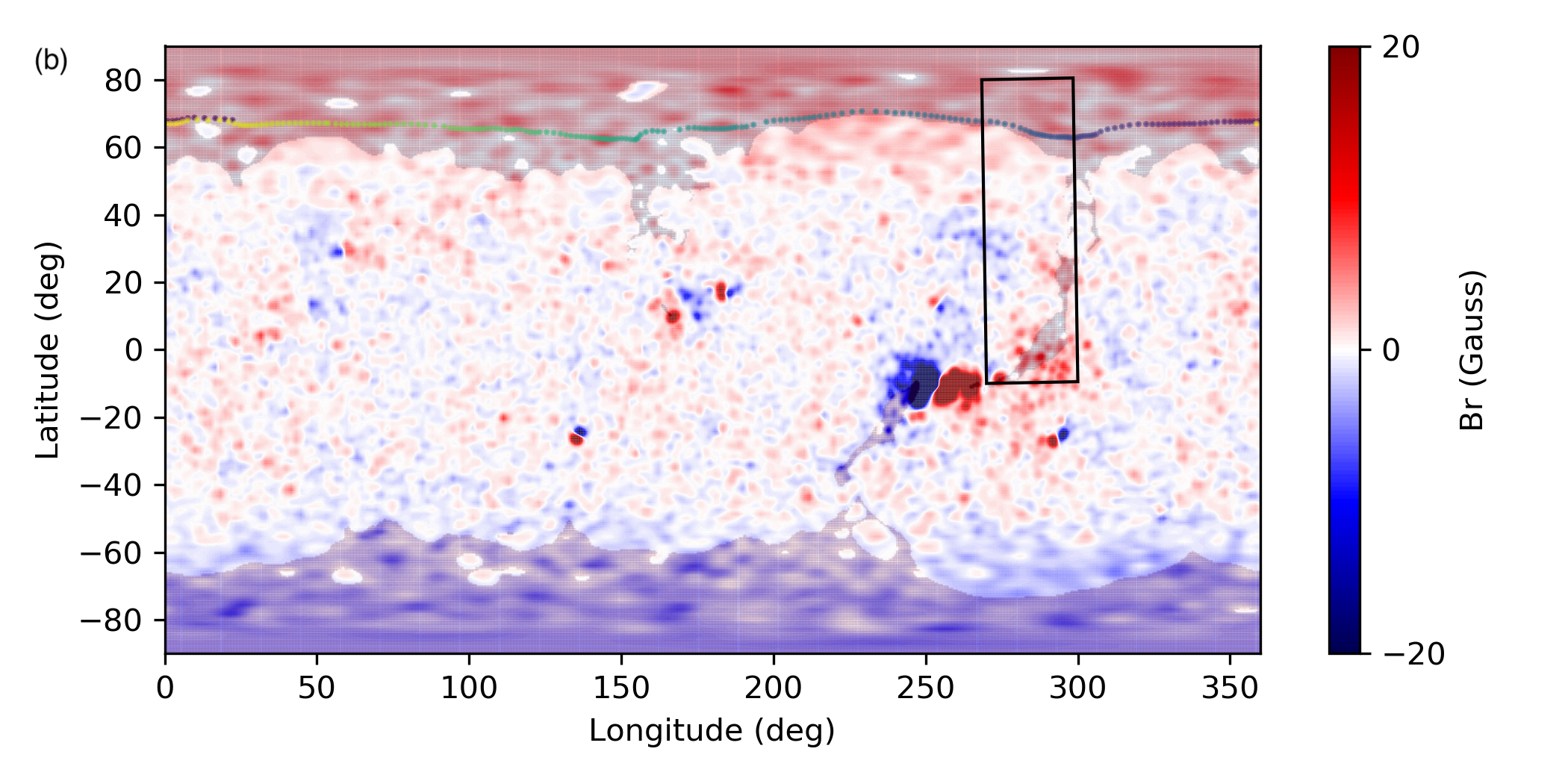}
\caption{(a) Magnetic field lines from MAS model for CR 1913; (b) The $B_{r}$ distribution at the solar surface. 
The dotted lines indicate the rooted position of streamlines at the solar surface and 
the dark shadow regions are for open field regions. The rectangle shows the pseudo-streamer region }
\label{fig: fieldlines}
\end{figure*}

The 3D MHD model takes in inputs for the radial magnetic field as a function
of colatitude and colongitude
governed by available full-Sun magnetogram data
and the coronal base temperature and density. The simulation then outputs a range of variables such as the plasma temperature, density, and velocity of the solar wind in steady-state \citep{2008SSRv..136..565A}.
By matching the observation dates with CR 1913, we were able to compare winds speeds calculated at the 20 $R_\odot$ boundary with those measured by SWICS\@.
Assuming the radial component of the velocity
profile in the MAS code dominates, Figure (\ref{fig: velocity}) shows the bulk flow velocity calculated by the MAS code at the 20 $R_\odot$ boundary (solid curve) and compares it with the measured ionic velocities measured by SWICS (dashed curve). The rectangular bar indicates when the pseudo-streamer appeared in the MHD model. The observed velocity profile was the steadiest between the dates of August 23 to September 9. After September 10, the observed wind speed continually declines for the rest of CR 1913. This decline is not captured by the MAS model. Therefore, we will focus the period from August 23 to September 9 in this comparison. We found that the SWICS observation shows a prominence of very fast solar wind (upwards of 700 \kms) while the MAS model
is computing values in the range of 460 \kms\ up to 540 \kms\@. This is likely indicative of an acceleration mechanism not accounted for by the MAS model. As a check, we plot the solar magnetic field configuration in Figure (\ref{fig: fieldlines}) and the observational region (in panel b) shows
that the field configuration for the pseudo-streamer greatly changes in the 
latitudinal direction. A clear feature appearing in both the model and in situ observations is that the velocity profiles show a local minimum plasma speed around this pseudo-streamer except the model-predicted velocity at 20 $R_\odot$ is still lower than Ulysses/SWICS measurements at $\sim$ 4 AU which evidences an underpredicted velocity profile by the MHD model. Another possibility is that there is acceleration beyond 20 $R_\odot$. The fact of highly-ionized plasma occupying slow solar wind streams is also confirmed by our analyses, but again the MHD wind speeds do not agree with the in situ measurements of SWICS. 

In MHD modeling cases where the plasma is leaving the corona slower than in reality, the plasma will have more time for the charge states to evolve. The observed charge states could have frozen in at higher heights. For example, if the plasma is dropping in temperature as it moves away from the Sun, the charge states may correspond to lower temperatures than observed by Ulysses. This may contribute as one of the sources of errors in the following comparison between NEI calculations and Ulysses/SWICS observations.

\begin{figure*}
\includegraphics[width=\textwidth]{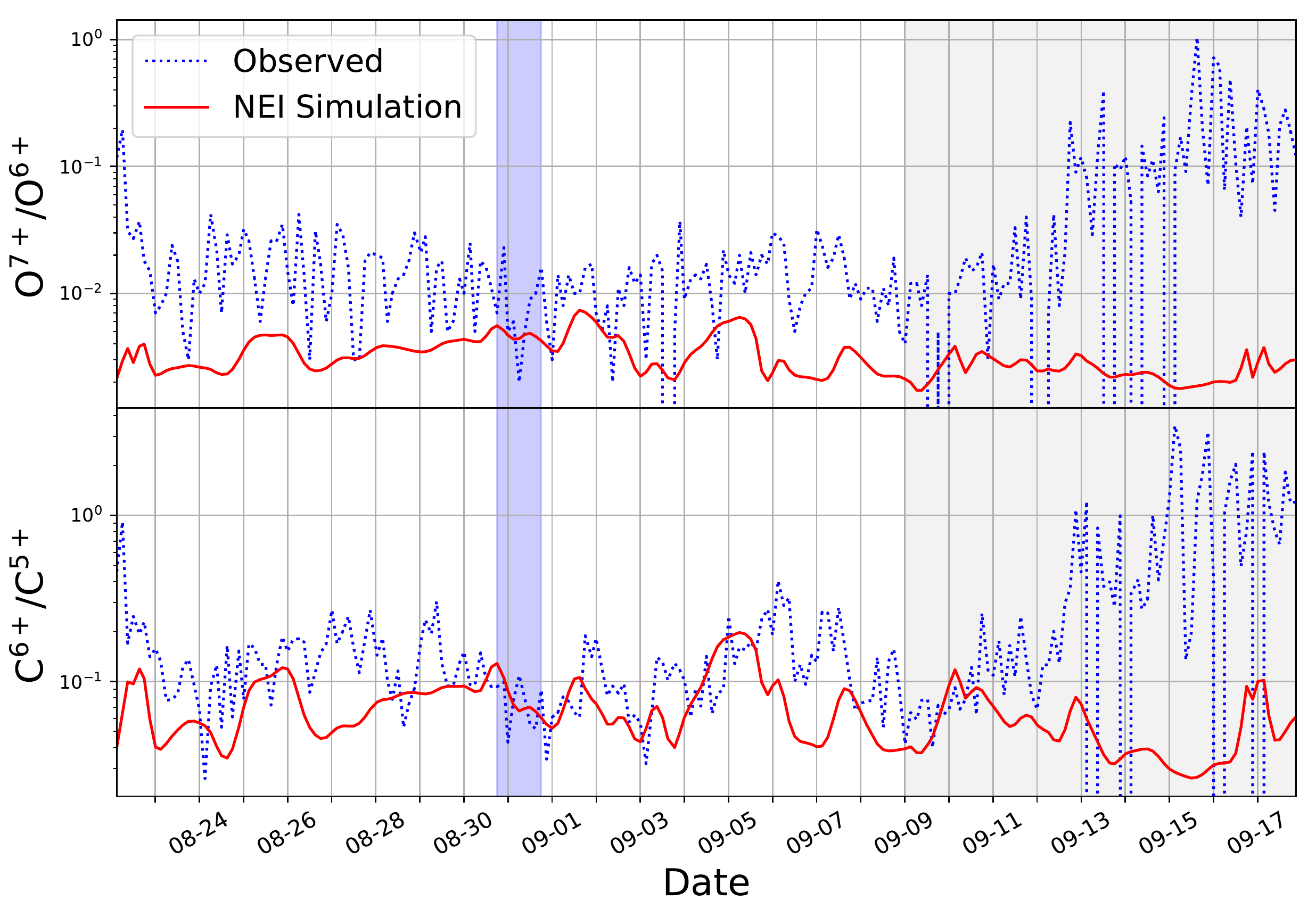}
\caption{The abundance ratios of solar wind ions measured by the SWICS instrument compared
with the calculated values of our NEI model along with the labeled appearance of the pseudo-streamer. The grey area indicates where SWICS began to observe a decline in solar wind velocity which agrees with the anti-correlated nature of high ionization
in the slower wind streams. }
\label{fig: abund}
\end{figure*}

\subsection{Charge State and Abundance Ratio Comparison}\label{sec: abund_compare}
\begin{figure*}[t]
\includegraphics[width=\textwidth]{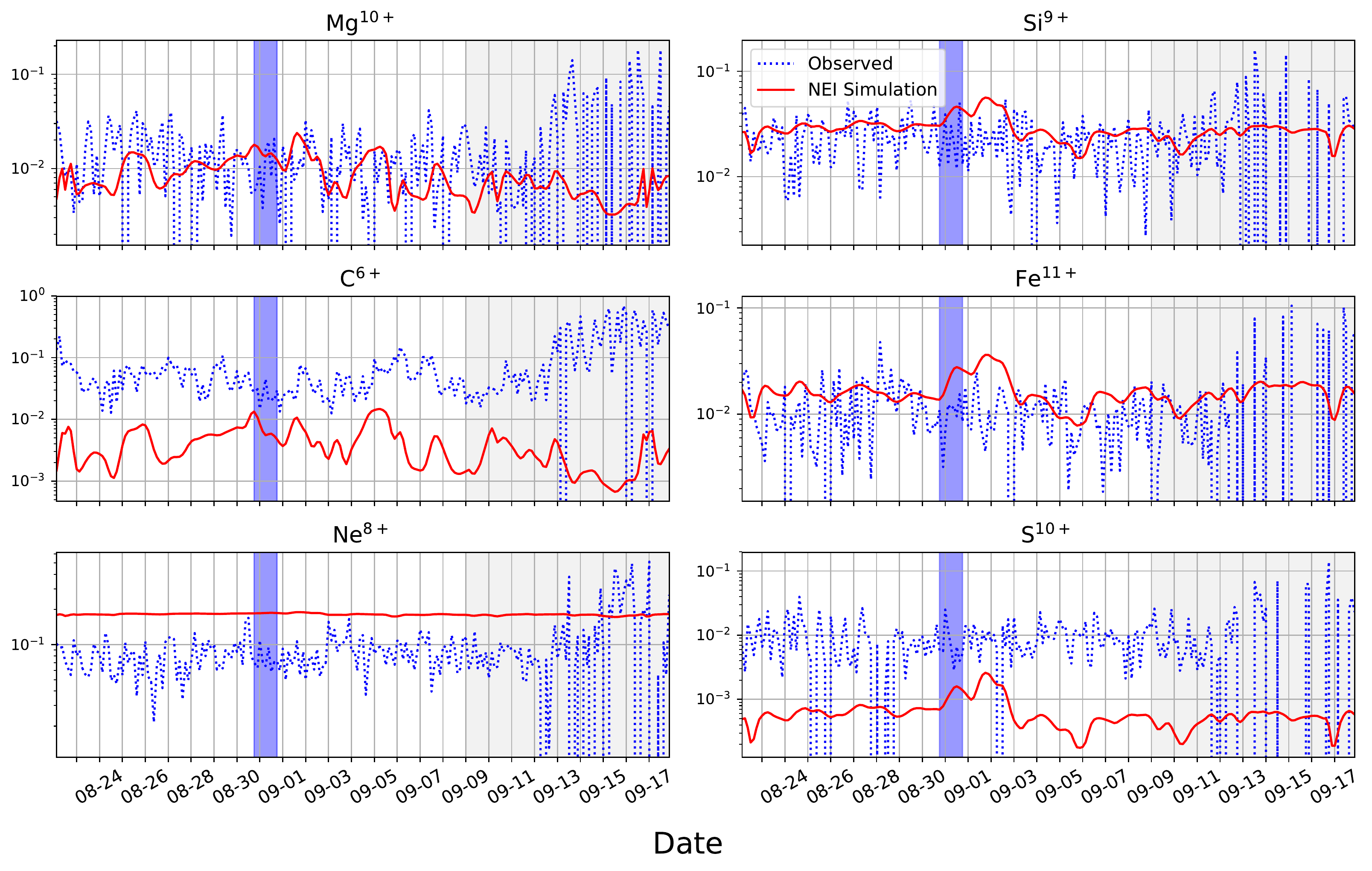}
\caption{The ionic densities relative to O$^{6+}$ for the main solar wind ions measured by SWICS\@. The observation (blue) is superimposed with the model (red) where the model is in reasonable agreement for a few of the ions in question. The blue rectangle indicates the pseudo-streamer observation.
}
\label{fig: density}
\end{figure*}

\begin{figure*}
\includegraphics[width=\textwidth]{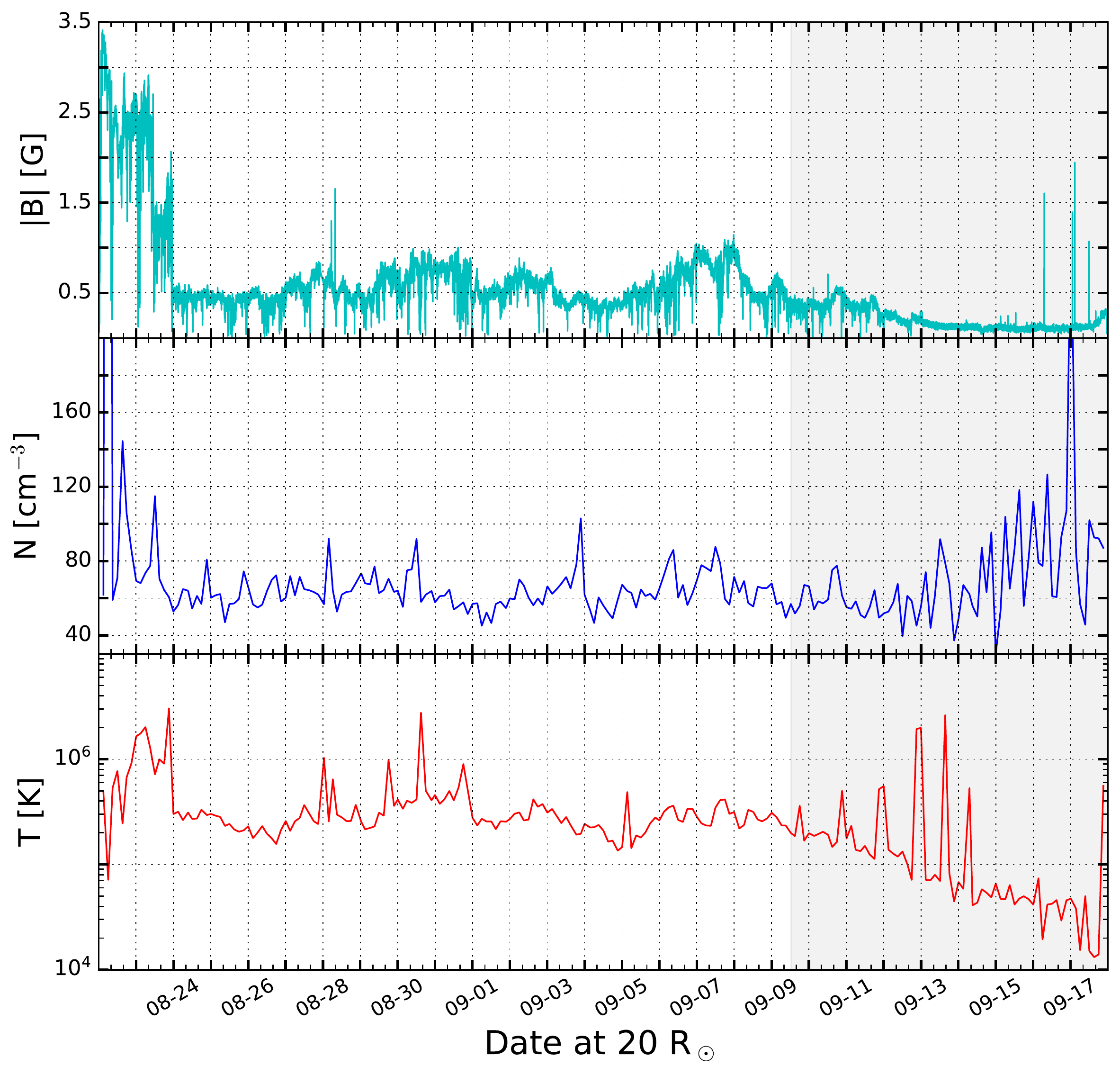}
\caption{Shown above are the magnetometer data, $\alpha$-particle density,
and $\alpha$-particle temperature measured by SWICS\@. The grey region indicates
when the observed winds begin to decelerate. This important distinction is made
for the peculiar result shown in Figure (\ref{fig: abund}) .}
\label{fig: mag}
\end{figure*}

In this section, we calculated the final non-equilibrium ionic fractions at the 20 $R_\odot$ boundary based on temperature and density values predicted by the MAS model. We used ionization/recombination tables assuming the Maxwellian 
distribution throughout the NEI  simulation. Although non-Maxwellian electron distribution functions exist in the ion-forming region \citep{2003AIPC..679..249E}, we assume these effects to be negligible in reference to our calculations.  The O$^{7+}$/O$^{6+}$ and 
C$^{6+}$/C$^{5+}$ ratios measured by the SWICS instrument were 
compared with the results of the NEI simulation.
Observed in situ abundance ratios are shown in Figure (\ref{fig: abund}) where we have superimposed the NEI calculated fractions on top of the SWICS measurements. The blue strip indicates when the pseudo-streamer outflow would have made it to SWICS, and the grey region indicates
the start of the decline in observed solar wind speed as seen in Figure (\ref{fig: velocity}). Figure (\ref{fig: density}) shows the relative densities of each measured ion versus the NEI calculated ratios. The simulation lies within reasonable agreement for the Fe$^{11+}$, Mg$^{10+}$, Si$^{9+}$, but does not match well with the other ions in question. An interesting observation is that 
the region of solar wind decline is highly correlated to a higher degree of ionization for all observed solar wind ions which would suggest an increase in temperature and magnetic field strength, but Figure (\ref{fig: mag}) contradicts that notion. The O$^{7+}$/O$^{6+}$ ratio is a proxy for the degree of ionization in solar plasma, and the observed ratio increase towards the end of the WSM (Figure \ref{fig: abund})  might be attributed to the plasma having more time to ionize as it moved away from the Sun. 

\subsection{Charge State Freeze-In Heights}\label{sec: freeze_heights}
In wanting to understand the heavy ion distribution in reference to their
expansion into the solar wind, we measured the ``freeze-in'' heights
of each solar wind ion 
from the model prediction.
\begin{figure*}[t]
\includegraphics[width=\textwidth]{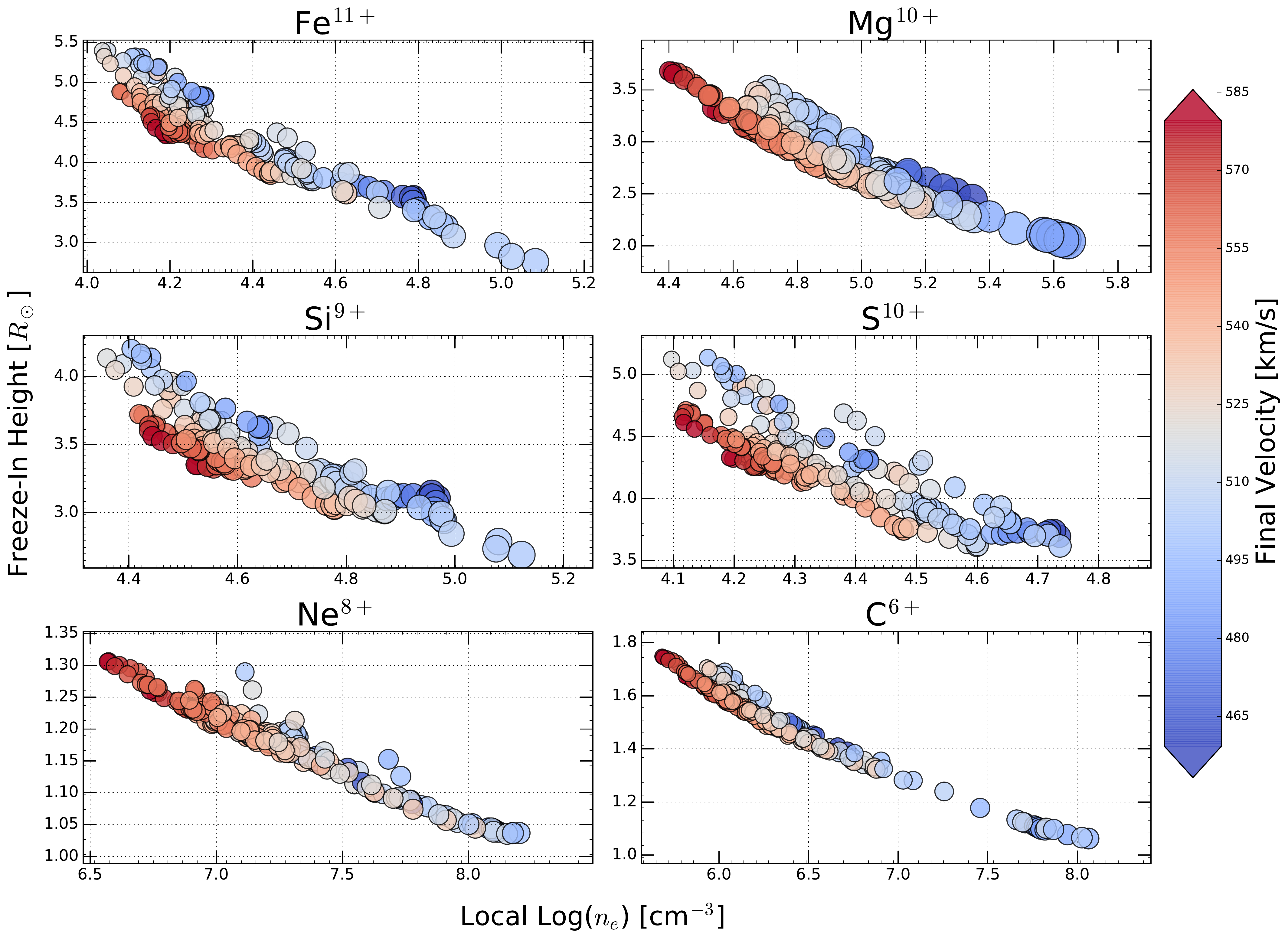}
\caption{Freeze-in heights obtained from the MHD model for the respective streamline
traces corresponding to Ulysses/SWICS observational data points. The horizontal-axis
indicates the local density when a particular ion freezed in, and the color
indicates the final plasma flow velocity at 20$R_\odot$. 
}
\label{fig: heights}
\end{figure*}
Measurements of the ions at freeze-in potentially yield information about the solar plasma conditions at the coronal base \citep{1983ApJ...275..354O}. 
For each position corresponding to the date of Ulysses/SWICS observations, we monitored the constant NEI states along each streamline and found the freeze-in height for each ion.
Figure (\ref{fig: heights}) shows
the heights at which each ion's ionization state becomes fixed with respect
to both the final velocity at 20 $R_\odot$ and the local freeze-in 
density along the flux tube. We can see a trend of ions
freezing in at higher heights based on the magnitude of their velocity profiles.
This corresponds to the fact that plasma that is moving quickly from the solar surface takes comparatively longer to equilibrate. 
It is also clear that freeze-in heights dramatically drop as the local density increases because the ionization/recombination time-scales become short in dense plasma environments. For relatively low-temperature ions, such as C$^{6+}$ and Ne${^8+}$, the ions generally freeze-in lower altitudes (1.8 and 1.3$R_\odot$) even for the high speed flows with velocity up to ${\sim}$ 600 \kms\@. On the other hand, the high-temperature ions (Si${^{9+}}$, S${^{10+}}$, Fe${^{11+}}$) show much wider freeze-in height ranges depending on the flow speed. Along with the freeze-in heights-density profiles in Figure (\ref{fig: heights}), we noticed that our analysis does not show the higher density ranges ($> {10^6}$ cm${^{-3}}$) for Fe${^{11+}}$ due to the limited sampling points from the current MHD model. It will be interesting to investigate the freeze-in heights in more dense plasma cases in the future. 
Moreover, we have included the freeze-in heights for Fe$^{10+}$ and Fe$^{13+}$ in one figure (Figure \ref{fig: iron_freeze_in}) to discuss the variation of freeze-in heights with charge states. The overall feature of Figure \ref{fig: iron_freeze_in} is that Fe$^{10+}$ ions freeze in at lower heights first, and Fe$^{13+}$ fraction becomes fixed at the higher altitudes. This tendency is consistent with recently solar wind observations \citep[e.g.,][]{2018ApJ...859..155B}.
\citet{2018ApJ...859..155B} provided reliable measurements for the freeze-in heights of the aforementioned ions for coronal holes. They measured the freeze-in distances to be in the range of
1.4 to 2 $R_\odot$ for Fe$^{10+}$ and from 1.5 to 2.2 $R_\odot$ for Fe$^{13+}$ in 
open field streamer regions during the 2015 March 20 total solar eclipse. In here, we compute a wider range of freeze-in heights in our model, showing how the dynamics of different coronal features can greatly affect when and where ionization states become fixed. Note that in the right corner of the Fe$^{10+}$ panel in Figure (\ref{fig: iron_freeze_in}), the freeze-in heights match closely to those found in \citet{2018ApJ...859..155B} while the Fe$^{13+}$ distances are a bit higher partly due to the absence of higher density sampling streamlines from the current MHD model.
\begin{figure*}[t]
    \includegraphics[width=\textwidth]{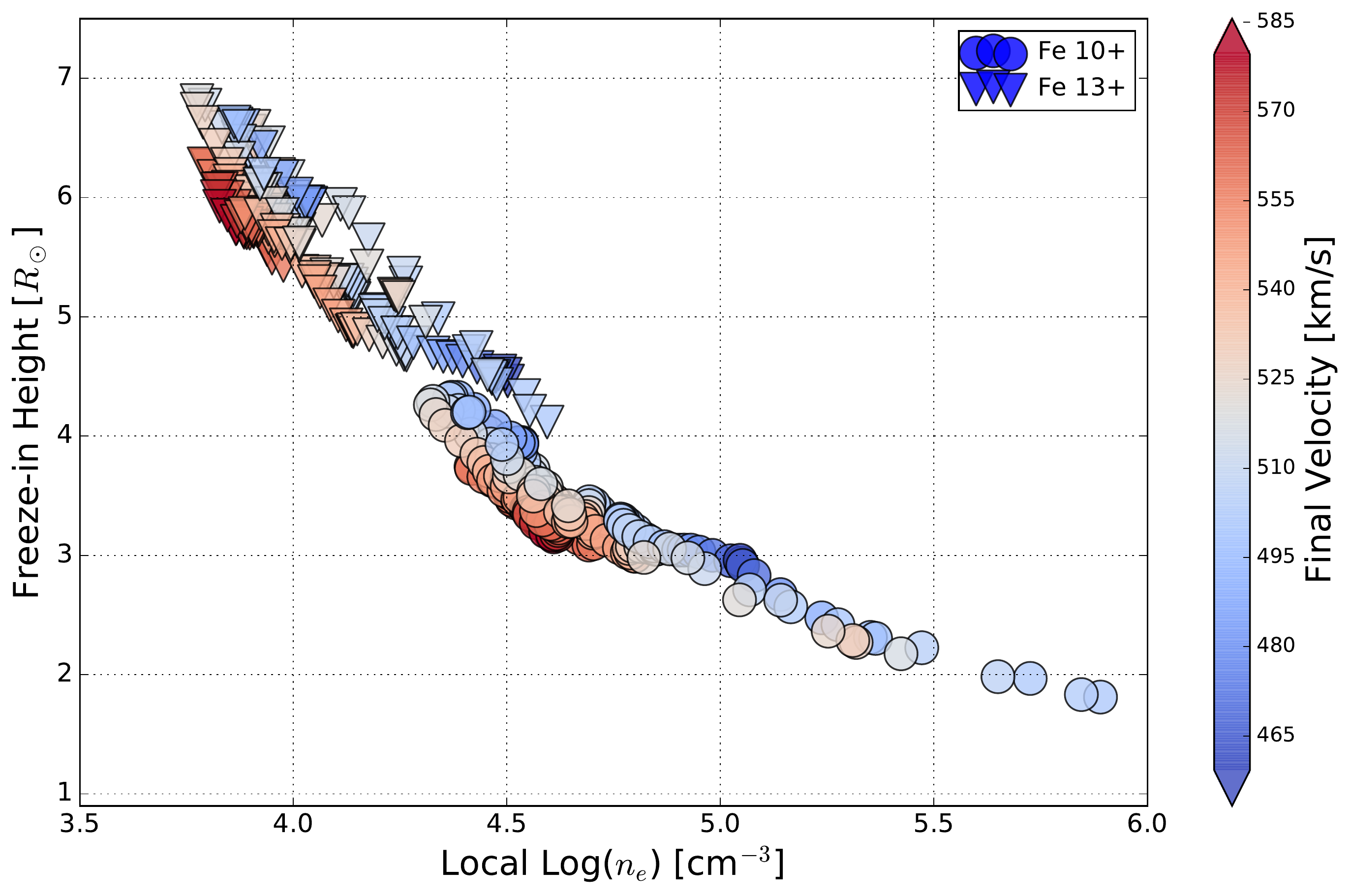}
    \caption{The freeze-in distances vary with local density and final plasma flow speed for Fe$^{10+}$ and
    Fe$^{13+}$ ions calculated from the MHD model. The sampling points are chosen to be the same as in Figure \ref{fig: heights}.
    }
    \label{fig: iron_freeze_in}
\end{figure*}
Overall, the ion density to freeze-in height relationship corresponds well with the notion that local density falls with the height. Additionally, Figures(\ref{fig: heights} \& \ref{fig: iron_freeze_in}) both show that faster moving ions freeze in at higher heights. What is not fully understood is the order-of-magnitude differences between the local densities. Nonetheless,
these results are useful for one seeking to provide constraints on freeze-in
distances and local densities given the outflow velocities of a wind near the northern edge variability band as described in \citet{1997GeoRL..24..309G}. For instance, one looking to estimate the freeze-in distance
for Mg$^{10+}$ that is moving at a velocity greater than 570 \kms\ can expect to 
see the ionization state become fixed beyond a distance of 3.5 $R_\odot$.

\section{Summary} \label{sec:discussion}
In this analysis, we used NEI calculations 
to compare model predicted charge state populations with those observed by the
Ulysses/SWICS instrument during the WSM interval CR 1913 (1996 August 22 to September 18).
We calculated charge states of abundant elements in solar wind by using the plasma evolution obtained from the MAS model up to 20 $R_\odot$.
We then assume a steady velocity thereafter until the plasma reaches the SWICS instrument located at about 4.25 AU during this WSM interval. We compared the modeling plasma velocity at 20 $R_\odot$ versus the bulk flow velocities measured by Ulysses/SWICS\@. The discrepancies shown in Figure (\ref{fig: velocity}) might be attributed to the MHD model not accounting for acceleration within 20$R_\odot$ or that the plasma accelerates beyond the modeling boundary. 
For a discussion of solar wind acceleration mechanisms beyond 20 $R_\odot$, we refer the reader to work such as those by \citet{McGregor:2011}.

We compared densities (relative to O$^{6+}$) of the SWICS ions (C, O, Ne, Mg, S, Si, Fe) with our NEI calculations and found that Fe$^{11+}$, Mg$^{10+}$, and Si$^{9+}$ ratios matched observation much better than S$^{10+}$, Ne$^{8+}$, and
C$^{6+}$. We then measured the modeled O$^{7+}$/O$^{6+}$ and C$^{6+}$/C$^{5+}$ ratios with the in situ measurements and found reasonable agreement up until the drastic change in ion population that we assume was due to the interaction region ionizing the plasma beyond 20 $R_\odot$. To check for any sign of solar energetic events, we plotted the observed magnetic field strength, $\alpha$-particle density, and temperature in Figure (\ref{fig: mag}). We might infer that the abrupt changes in these characteristics starting from the grey region in the plot provide more evidence of a heliospheric shift in the plasma conditions that was not captured by the MHD simulation. The velocity profile transitioning to the slow wind regime implies a mixing of plasma domains \citep{1994GeoRL..21.2063P, 1997GeoRL..24..309G, 1999SSRv...89...21G} that might explain the large discrepancy between our model and observations. To be certain, we searched all available data for signs of solar energetic events during this WSM interval to no avail. In such a search, one might check for the presence a high-energy event via the Large Angle and Spectrometric Coronagraph (LASCO) aboard SOHO, but no such event was present in the data.

We calculated the freeze-in heights --- the point at which ionization temperature is constant ---  for ions while relating this quantity to both the local plasma density at freeze-in as well as the plasma final velocity at 20$R_\odot$. This method can potentially be useful for observers wanting to know reference values for particular ions given initial conditions of the plasma. 
The results we found for Fe$^{10+}$ and Fe$^{13+}$ offer a wider range of freeze-in values than those reported  by \citet{2018ApJ...859..155B}. Our calculated freeze-in heights ranges for Fe$^{10+}$ (4.5 to 7.5 $R_\odot$) and Fe $^{13+}$ (4.0 to 7.0 $R_\odot$) were greater than the combined ranges for both Fe$^{10+}$ (1.4 to 2 $R_\odot$) and
Fe$^{13+}$ (1.5 to 2.2 $R_\odot$) examined in the \citet{2018ApJ...859..155B} work --- with the effects of systematic error taken into account --- but we emphasize that this discrepancy can be attributed to a wide range of uncertainties in MHD simulations, such as inaccurate acceleration within 20$R_\odot$. It is also likely to be an inherent fundamental difference between the magnetic field regions at which our two observations occur. Although under different magnetic field configurations, we cite the work of \citet{2018ApJ...859..155B} for comparison as a means of demonstrating the curious physical effects at play between different streamer regions. Future studies could shine more light on how impactful both coronal and ambient magnetic field configurations affect the freeze-in distance for many solar wind ions.

NEI calculations are highly sensitive
to electron density and temperature evolutionary history.
Differences between the true solar wind velocity 
and the velocity given by the simulation will contribute to discrepancies 
between the predicted and observed charge state distributions.
Although the discrepancy of ion populations
between observations and models in the 
past have been hypothesized to be due to non-thermal electrons
\citep{2003AIPC..679..249E}, the differences between non-Maxwellian versus
Maxwellian distributions in terms of the degree of ionization in the solar
wind are negligibly small \citep{2007ApJ...660.1642L}. This is worth exploring in the future. 
Since the simulation domain of this MHD model is up to 20 $R_\odot$, it cannot reveal the plasma movement and directly obtain the velocity prediction at Ulysses' orbit. Note that our driving assumption throughout this procedure was that
the plasma velocity remains constant from 20 $R_\odot$ onward. The error in not accounting for acceleration beyond the 20 $R_\odot$ boundary is beyond the scope of this paper. It is possible that the solar wind accelerated in the interplanetary medium by means of wave-particle interactions such as shocks, but signs of shock-producing events such as coronal mass ejections (CMEs) were not detected by available coronagraphs. However, the model's velocity profile provides sufficient information on the plasma before acceleration. 

In essence, the variation in solar wind speed can come from a variety
of sources and tracing a parcel of plasma back to the source is
quite difficult at 1 AU let alone 4 AU\@. Future work would be to implement
different electron distributions in the NEI model such as the kappa distribution which
accounts for non-Maxwellian suprathermal tails. 
Velocity calculations 
in the MAS code may also be an important step towards improving the NEI models
since they rely so heavily on the magnetic field configuration. 
The large discrepancy between the observed velocity profile in the MAS model tells us
that either the MHD code is largely underestimating the velocity of the plasma outflow, 
we were incorrect in assuming the velocity was purely radially
outward at the 20 $R_\odot$ boundary, or a combination of both effects. A more sophisticated
approach of back tracing a plasma flow from 4 AU to the Sun 
would be to use an inverse MHD mapping as discussed in \citet{Riley:1999}. This way, the plasma
parcel is accounted for throughout its entire journey and the appropriate acceleration mechanisms 
can be properly assessed throughout the trajectories.
We acknowledge that in our report there are systematic errors that might have produced the
differing effects, but such errors are not easily quantifiable. One way to smooth
the systematic error in future work would be to expand our observation region to account for solar latitudes that might match more closely to observation.

This work was primarily supported by NSF SHINE Grant AGS-1723313 to the Smithsonian Astrophysical Observatory (SAO) and a Smithsonian Institution Scholarly Studies grant.
C.S.\ is supported by NSF grant AST-1735525, and NASA grants 80NSSC20K1318, 80NSSC19K0853, and 80NSSC18K1129 to SAO.
N.M.\ acknowledges partial support by NSF grant 1931388, NASA grants 80NSSC19K0853 and 80NSSC20K0174, and NASA contract NNM07AB07C to SAO\@.\\

\large{\textit{Facilities}:} Ulysses, SOHO\\

\large{\textit{Software}:}
    MAS, 
    NEI \citep{Shen2015b}, 
    HelioPy \citep{heliopy_0_5_1}, 
    Astropy \citep{2018AJ....156..123A}, 
    PlasmaPy \citep{plasmapy_0_1_1, plasmapy_community_2018_1238132},
    CHIANTI, v8.0.7, is a collaborative project involving George Mason University, the University of Michigan (USA), University of Cambridge (UK) and NASA Goddard Space Flight Center (USA).

\begin{acks}
The authors acknowledge helpful discussions with W.\ Barnes, J.\ Raymond, K.\ Reeves, D.\ Stansby and M.\ Stevens; and thank PSI for providing the MAS simulation.
\end{acks}




\bibliographystyle{spr-mp-sola}
\bibliography{refs}  

\IfFileExists{\jobname.bbl}{} {\typeout{}
\typeout{****************************************************}
\typeout{****************************************************}
\typeout{** Please run "bibtex \jobname" to obtain} \typeout{**
the bibliography and then re-run LaTeX} \typeout{** twice to fix
the references !}
\typeout{****************************************************}
\typeout{****************************************************}
\typeout{}}



\end{article} 

\end{document}